\documentstyle[12pt,epsfig]{article}
 1
\def\as{\alpha_{\rm S}}
\def\dd{\partial}

\def\T{{\cal T}}
\def\pom{{I\!\!P}}
\def\alphapom{{\alpha_{\pom}}}
\def\regg{{I\!\!R}}
\def\alpharegg{{\alpha_{\regg}}}

\def\precent{{\%\ }}
\def\citenum#1{{\def\@cite##1##2{##1}\cite{#1}}}
\def\citea#1{\@cite{#1}{}}

\def\lsim{\;\raisebox{-.4ex}{\rlap{$\sim$}} \raisebox{.4ex}{$<$}\;}

\def\as{\alpha_{\rm S}}

\def\T{{\cal T}}

\def\gev{{\,\mbox{GeV}  }}
\def\gev2{{\,\mbox{GeV}^2}}
\def\to{\rightarrow}

\def\a{\alpha}

\def\l{\lambda}

\def\s{\sigma}

\def\({\left(}
\def\){\right)}

\def\citenum#1{{\def\@cite##1##2{##1}\cite{#1}}}
\def\citea#1{\@cite{#1}{}}

\def\l1vt{\vec{l_{1\perp}}}

\def\bt{b_{\perp}}

\def\bkt{{\mathbf k_t}}
\def\bkt2{{\mathbf k^2_t}}
\def\bt2{$b^2_t$}

\def\jol1{$J_0(\,l_{1\perp}\,r_{\perp}\,)$}

\def\citea#1{\@cite{#1}{}}

\def\beq{\begin{equation}}
\def\eeq{\end{equation}}
\def\bea{\begin{eqnarray}}
\def\eea{\end{eqnarray}}

\def\eq#1{{\mbox{Eq.\hspace{1mm}(\ref{#1})}}}
\def\eqs#1#2{{\mbox{Eqs.\hspace{1mm}(\ref{#1})--(\ref{#2})}}}
\def\fig#1{{\mbox{Fig.\hspace{1mm}\ref{#1}}}}

\def\ds{\displaystyle}
\def\scrbox#1{\mbox{\scriptsize #1}}

\def\bbbz{{\mathchoice {\hbox{$\sf\textstyle Z\kern-0.4em Z$}}
{\hbox{$\sf\textstyle Z\kern-0.4em Z$}}
{\hbox{$\sf\scriptstyle Z\kern-0.3em Z$}}
{\hbox{$\sf\scriptscriptstyle Z\kern-0.2em Z$}}}}
\def\npb#1#2#3{    {\it Nucl. Phys. }{\bf B#1} (19#2) #3}
\def\plb#1#2#3{    {\it Phys. Lett. }{\bf B#1} (19#2) #3}
\def\prd#1#2#3{    {\it Phys. Rev. }{\bf D#1} (19#2) #3}

\def\zpc#1#2#3{    {\it Z. Phys. }{\bf C#1} (19#2) #3}

\def\jetp#1#2#3{   {\it Sov. Phys. }{JETP }{\bf #1} (19#2) #3}

\def\epj#1#2#3{    {\it Eur. Phys. J.}{\bf #1} (19#2) #3}

\relax
\begin{document}
%
%
\newcommand{\sgp}{\sigma(\gamma^* p)}
\newcommand{\tm}{\tilde{M}}
\newcommand{\sigs}{\sigma^{\scrbox{soft}}}
\newcommand{\sigh}{\sigma^{\scrbox{hard}}}
\newcommand{\sigst}{\sigma^{\scrbox{soft}}_{\scrbox{T}}}
\newcommand{\sight}{\sigma^{\scrbox{hard}}_{\scrbox{T}}}
\newcommand{\sightQQ}{\sigma^{\scrbox{hard}}_{\scrbox{T},\bar{Q}Q}}
\newcommand{\sigsl}{\sigma^{\scrbox{soft}}_{\scrbox{L}}}
\newcommand{\sighl}{\sigma^{\scrbox{hard}}_{\scrbox{L}}}
\newcommand{\sighlQQ}{\sigma^{\scrbox{hard}}_{\scrbox{L},\bar{Q}Q}}
\newcommand{\sigl}{\sigma_{\scrbox{L}}}
\newcommand{\sigt}{\sigma_{\scrbox{T}}}
\newcommand{\lamlam}{{\lambda\lambda'}}
\newcommand{\qbarq}{\bar{q}q}
\newcommand{\aem}{\alpha_{\scrbox{em}}}
\newcommand{\mt}{\tilde M}
\newcommand{\lt}{\ell_{\perp}}
\newcommand{\lts}{\ell^2_{\perp}}
\newcommand{\kt}{{k_{\perp}}}
\newcommand{\kts}{{k^2_{\perp}}}

\newcommand{\softformula}[1]{\frac{\a_{em}}{3\pi}\int_0^{M_0^2} 
                          \frac{R(M^2) #1 dM^2}{\left( Q^2+M^2\right)^2}}
\newcommand{\dlformula}{A ( \frac{1}{x_M})^{
                         \alphapom(0)\,-\,1}\,\,+\,\,B\,\, (
                         \frac{1}{x_M})^{\alpharegg(0)\,-\,1}}
\begin{titlepage}
\noindent
\begin{flushright}
\parbox[t]{10em}{
TAUP 2573/99\\
{\today}\\
  } 
\end{flushright}
\vspace{1cm}
\begin{center}
{\Large \bf  Total ${\mathbf \gamma^* p}$ Cross Section}
 \\[4ex]

{\large E. ~G O T S M A N${}^{1)}$,\,\,\,\,\,\,\, E. ~L E V I
N${}^{2)}$,}

{\large
  U. ~M A O R${}^{3)}$\,\,\, and \,\,\,E. ~N A F T A L I$^{4)}$}
\footnotetext{$^{1)}$ Email: gotsman@post.tau.ac.il .}
\footnotetext{$^{2)}$ Email: leving@post.tau.ac.il .}
\footnotetext{$^{3)}$ Email: maor@post.tau.ac.il .}
\footnotetext{$^{4)}$ Email: erann@post.tau.ac.il .}
\\[4.5ex]
{\it  School of Physics and Astronomy}\\
{\it Raymond and Beverly Sackler Faculty of Exact Science}\\
{\it Tel Aviv University, Tel Aviv, 69978, ISRAEL}\\[4.5ex]

\end{center}
~\,\,
\vspace{1cm}

{\samepage
{\large \bf Abstract:}

This paper shows that the approach based on Gribov's ideas for the
photon-proton interaction, is able to describe the experimental data 
 over a
wide range of photon virtualities $Q^2 = 0 - 100\,GeV^2$ and 
 energies
$\sqrt{s} = W = 3 - 300\,GeV$. A simple model is suggested which provides
a quantitative way of describing the matching between short and long
distances ( between ``soft" and ``hard" processes ) in photon-proton
interaction at high energy. The main results of our analysis are: (i) 
the values of the separation parameters differentiating  
the ``soft" and ``hard"
interactions are determined; (ii) the additive quark model can be used to
calculate
 the
``soft" contribution to the photon - proton interaction; (iii)  a 
good
description of the ratio $\sigma_L/\sigma_T$ is obtained ; and (iv)  a
considerable
contribution of the ``soft" process at large $Q^2$, and of
``hard" processes  at small $Q^2$ was found. 

}
\end{titlepage}

\section{Introduction}
The rich and high precision data  on deep inelastic $e p $ scattering
at HERA \cite{H1} \cite{ZEUS}, covering both low and high $Q^2$ regions,
lead to a theoretical problem of matching the non-perturbative (``soft")
and perturbative (``hard ") QCD  domains. This challenging problem has
been under close investigation over the past two decades, starting from
the pioneering paper of Gribov \cite{GRIBOV} ( see also \cite{GVD} ).

\begin{figure}[tbp]
\begin{center}
  \epsfig{file=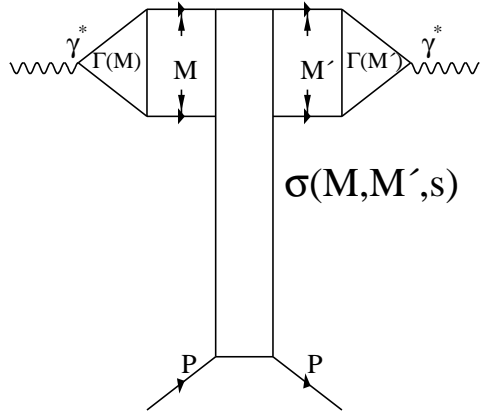,width=80mm}
  \caption[]{\parbox[t]{
             0.80\textwidth}{\small \it The generalized Gribov's formula.}}
\end{center}
\label{fig1}
\end{figure}

Based on Gribov's general approach, one can interpret two 
 time sequence of the 
 $\gamma^* p$ interaction (see \fig{fig1}):
\begin{enumerate}
\item\,\,\,First the $\gamma^*$ fluctuates  into  a hadron system
(quark-antiquark
  pair to  the lowest order) well before the interaction with the
target;

\item\,\,\,Then the  converted quark-antiquark pair (or hadron
system)
 interacts with the target.

\end{enumerate}
These two stages are expressed explicitly in the double dispersion relation
suggested in Ref. \cite{GRIBOV}:

\begin{equation}\label{GRIBOVgen}
\sigma(\gamma^* N)=\frac{\a_{em}}{3\pi}\int
  \frac{\Gamma(M^2) dM^2}{(Q^2+M^2)}
    \sigma(M^2,M'^2,s)
  \frac{\Gamma(M'^2) dM'^2}{(Q^2+M'^2)},
\end{equation}
where $M$ and $M'$ are the invariant masses of the incoming and outgoing
quark-antiquark pairs, $\sigma(M^2,M'^2,s)$ is the cross section of a
$q \bar q $ interaction with the target, and the vertices $\Gamma^2(M^2)$ and
$\Gamma^2(M'^2)$ are given by  $\Gamma^2(M^2)=R(M^2)$, where $R(M^2)$ is the
ratio: 
\begin{equation}\label{Rdef}
R(M^2)=\frac{\sigma(e^+e^-\to \mbox{hadrons})}{\sigma(e^+e^-\to
\mu^+\mu^-)},
\end{equation}
which has beem measured experimentally.  
 For large masses 
($M,\, M' >> M_0$) we have  
$\,\Gamma(M^2\,\geq\,M^2_0)\times \Gamma(M'^2\,\geq\,M^2_0 )\,\,
\longrightarrow
\,\,R( M^2
\,\geq\,\,M^2_0 ) \,=\, 2$, where we assume that that the  number of
colours $N_c = 3 $. $M_0$ is a typical mass (a separation 
parameter), which is of the order
of $M_0 = 1  GeV $, determined directly from the experimental data
\cite{TABLE}.

\begin{figure}[tbp]
\begin{center}
  \epsfig{file=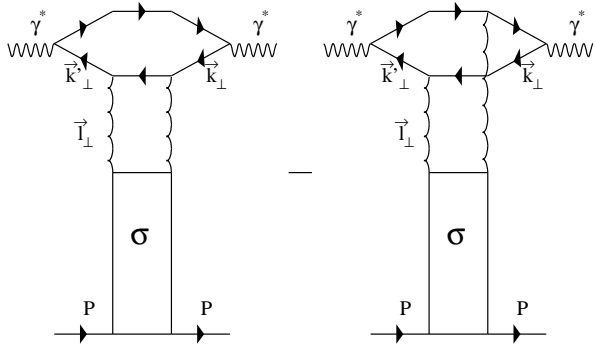,width=90mm}
  \caption[]{\parbox[t]{
             0.80\textwidth}{\small \it ``Hard" contribution to
$\sigma(M^2,M'^2,s)$ in perturbative QCD .}}
\end{center}
\label{twogluons}
\end{figure}

The key problem in all approaches utilizing  \eq{GRIBOVgen} is the
 description of the cross section $\sigma(M^2,M'^2,s)$. In this paper we
follow the approach suggested  in Ref.  \cite{GLMepj98}, which is based on
the ideas 
 of Badelek and Kwiecinski  \cite{BK}.  Below we list the
mains steps of this approach.
\begin{enumerate}
\item\,\,\,We introduce $M_0 \,\,\approx \,\,1\,GeV$ in the integrals over
$M$ and $M'$ which plays the role of a separation parameter.  For $M,\, M'
\,>\,M_0$ the quark - antiquark pair are produced at short distances (
$r_{\perp}\,\,\propto\,\frac{1}{M}\,\,<\,\,\frac{1}{M_0}$ ), while for $M,\,
M'\,<\,M_0$ the distance between quark and antiquark is too long
($r_{\perp}\,\,\propto\,\frac{1}{M}\,\,>\,\,\frac{1}{M_0} $), and we cannot
treat this $q \bar q $ - pair in pQCD. Actually, we cannot even describe the
produced hadron state as a $q \bar q $ - pair;

\item\,\,\,For $M,\,M'\,<\,M_0$ we use the Additive Quark Model
\cite{AQM} in which
\begin{eqnarray} 
\sigma(M^2,M'^2,s) &=&\sigma^{soft}_N (M^2,s)
\,\delta(M^2\,-\,M'^2) \nonumber\\ 
 &=& \left(\,\sigma_{tot}(qN
)\,+\,\sigma_{tot}( \bar q  N )\,\right)\,\delta(M^2\,-\,M'^2)
\label{SOFT}
\end{eqnarray}
The above assumption allows us to simplify the Gribov formula of
\eq{GRIBOVgen}:

\begin{equation}\label{GRIBOVeq}
\sigma(\gamma^* N)=\frac{\a_{em}}{3\pi}\int
     \frac{R(M^2) M^2 dM^2}{\left(
Q^2+M^2\right)^2}\sigma_{N}(M^2,s)\,\,;
\end{equation}

\item\,\,\,For $M,\, M' \,>\,M_0$ we consider the system with mass $M$
and/or $M'$ as a short distance quark - antiquark pair, and describe
its interaction with the target in pQCD (see \fig{twogluons}).

The exact formulae for $\sigma^{hard}(M^2,M'^2,s)$ is given  below (see
also Ref. \cite{GLMepj98}), but one can see from \fig{twogluons},  that
this
interaction can be expressed through the gluon structure function, and it is
{\em not} diagonal with respect to the masses, contrary to the ``soft"
interaction of a hadron system with small mass.

\item\,\,\,In principle, the short distance between the quark and antiquark
  leads to short distances in the gluon-nucleon interaction. However, the
  typical distance of the gluon interaction $r_G\,\,\propto\,1/l_{\perp}$
   (see Figs 1 and 2) is larger than the size of the 
  quark - antiquark pair
  and $l_{\perp} \,\,<\,\,\kt$. It turns out that the calculation with
  $M_0\,\, \sim\,\,1\,GeV$,   demands a new scale in the gluon-nucleon
  interaction.  We introduce it, assuming that the gluon structure function
  $xG(x,l^2_{\perp})$ behaves as
  $xG(x,\mu^2)\,\times\,\frac{l^2_{\perp}}{\mu^2}$ for
  $l^2_{\perp}\,\leq\,\mu^2$.  This means that we assume that the gluon-hadron
  total cross section is not equal to zero for long distances (in the ``soft"
  kinematic region). It should be stressed that we introduce two scales for
  soft nonperturbative interactions, distinguishing between quark and gluon
  interactions. The two scales appear naturally in our models for ``soft",
  nonperturbative interaction.  For example, in the AQM which we use for
  $\sigs$, these two scales are the size of a hadron (distance between
  the 
  quark and antiquark), and the size of the constituent quark which is
  related to gluon interaction scale.

\item\,\,\, As has been discussed, we use the AQM [see \eq{SOFT}] to calculate
$\sigs$. For  the  vector meson cross section the AQM
leads to

\begin{equation}\label{RCRSEC}
\s_{\scrbox{R}}\,\,=\frac{1}{2}\left(\,\s(\pi^+ p)\,\,+\,\,\s(\pi^-p)\right).
\end{equation}
For the pion-proton cross section, we use the Donnachie - Landshoff
Reggeon parameterization \cite{DL}, with an energy variable which is
appropriate for the interaction of a hadronic system of mass $M$ with the
target (see \cite{GLMepj98} for details).
\begin{equation}
\label{DL}
\sigma_{M^2N}(s) = \dlformula,
\end{equation}
with
\begin{eqnarray} \label{alphapom}
\alphapom  &=& 1.079 \nonumber \\
\alpharegg &=& 0.55 \nonumber \\
x_M        &=& \frac{M^2}{s} \nonumber \\
A          &=& 13.1\, \mbox{mb} \nonumber \\
B          &=& 41.08\, \mbox{mb}\, .
\end{eqnarray}
The values of $A$ and $B$ are chosen so that (\ref{DL}) is valid for the
$\rho$-proton interaction. Using (\ref{DL}) in Gribov formula
(\ref{GRIBOVeq}) we derive the soft transverse contribution to $\sgp$:
\begin{equation}\label{soft2}
   \sigst = \softformula{M^2}\left\{\dlformula\right\}.
\end{equation}

\end{enumerate}

It has been shown in \cite{GLMepj98} that  even though the simple
model
discussed above reproduces the main features of the experimental
data, it has several deficiencies which we wish to address: 
\begin{itemize}
\item The calculations of the ``soft'' cross section appeared 
to overestimate the
data, and to overcome this,  the soft contribution was multiplied by a
factor
 $\kappa =0.6$. There is no physical justification for
such a small value.
\item The contribution of a longitudinal polarized virtual photon to $\sgp$
      was neglected.
\item An old GRV'94 parameterization was used for the gluons distribution
      inside the proton. This gave an energy dependence of
      $\sgp$ which is steeper than the published data.
\end{itemize}

In the present paper we  reexamine these points developing a
formalism which also takes into account $\sigl$, the longitudinal part
of $\sgp$, for both the soft and the hard components. We find that the
contribution of $\sigl$ significantly improves the energy dependence of
$\sgp$, in a good agreement with the data for $x < 10^{-2}$.

A similar approach to photon - proton interaction was also developed in
 Ref.\cite{MRS} where quite different scales have been introduced. The goal
 of such an approach  (see Refs.\cite{GLMepj98},\cite{BK} and
\cite{MRS}) is
 to find parameters that separate the long distance (non-perturbative)
 and short distance (perturbative) interactions in QCD. A similar
 philosophy to ours is used in Ref.\cite{BW}, where an attempt was made to
 describe photon - proton interactions, assuming that the main contribution
 stems from short distances, and the long distance interaction appears as
  a result of the  shadowing corrections.

The description of $\sigma( \gamma^* p )$ can be achieved in quite a
  different way using the generalized vector dominance model\cite{GVD} or
  a Regge motivated fit of the experimental 
  data\cite{ALLM}\cite{ALLM97}. In these  
  approaches one looses the explicit connection with the microscopic
  theory, which makes futher theoretical interpretation of the results rather
  difficult.

\section{Description of the Model \label{secMODEL}}
\subsection{The Contribution of a Transverse Polarized Photon \label{secT}}
\begin{figure}[tbp]
\begin{center}
  \epsfig{file=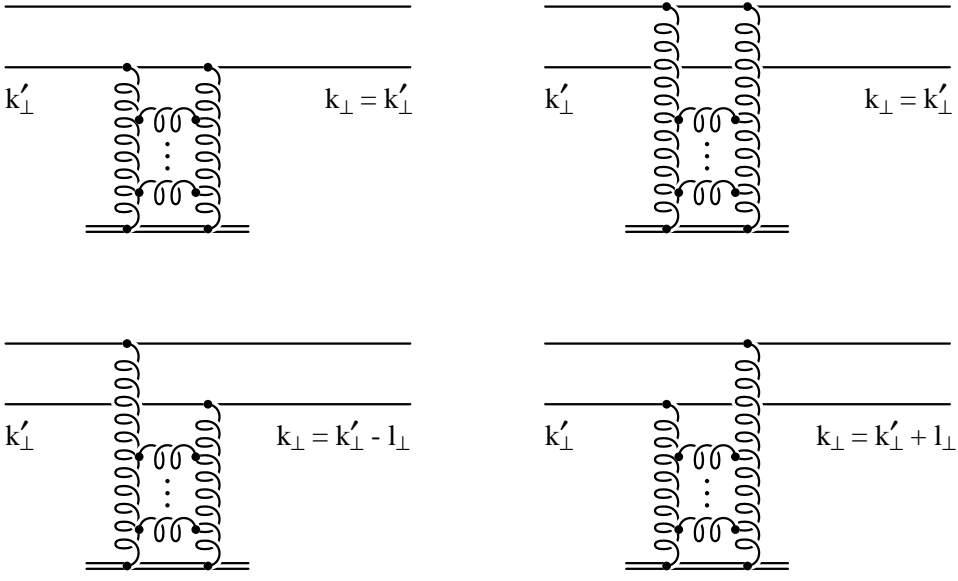,width=140mm}
  \caption[]{\parbox[t]{
             0.80\textwidth}{\small \it Diagrams which cntribute to the scattering
             of a $\qbarq$ pair off the target proton.}
             \label{fig3}}
\end{center}
\end{figure}
As we have discussed previously, the main  assumption of our model is
that
$\sigma(M^2,M'^2,s)$ in \eq{GRIBOVgen} can be calculated as \beq \label{S+H}
\sigma(M^2,M'^2,s)=\sigma^{soft}_N(M^2\,<\,M^2_0,s)\,\, \delta( M^2 - M'^2 )+
\sigma^{hard}(M^2\,>\,M^2_0,M'^2\,>\,M^2_0,s)\,\,.  \eeq In the previous
section we  have discussed 
the calculation of  $\sigs$. Now we present,
for completeness,  our formulae for $\sigh$ which have been given in
Ref.~\cite{GLMepj98}. The pQCD contribution for $\sigt$, within the framework
of the two gluon model, is illustrated in the four diagrams  shown in
Fig.~\ref{fig3}.  The production amplitude ${\cal M}_{\lamlam}$ of the
$\qbarq$, can be factorized into the wave function $\psi_\lamlam$ of the
$\qbarq$ system inside the virtual photon, and the amplitude $\T_\lamlam$ for
the scattering of the $\qbarq$ pair off the proton:
\begin{equation}\label{mlamlam1}
{\cal M}_\lamlam (\kt,z) = \sqrt{N_c}\int d^2 \kt'\int_0^1 dz'
                         \psi_\lamlam (\kt',z')\T_\lamlam (\kt',z';\kt,z)\, .
\end{equation}
The transition ampiltude $\T_\lamlam$ for the four contributions is:
\begin{eqnarray}\label{tlamlam}
\lefteqn{\T_\lamlam (\kt', \kt) =} & & \\
& &      i \frac{4\pi s}{2N_c} \int \frac{d^2 \lt}{\lt^4}
              \biggl [ 2 \delta (\mbox{\boldmath $k$}_{\perp}'    -
                                 \mbox{\boldmath $k$}_{\perp})    -
                         \delta (\mbox{\boldmath $k$}_{\perp}'    -
                                 \mbox{\boldmath $k$}_{\perp}     -
                                 \mbox{\boldmath $\ell$}_{\perp}) -
                         \delta (\mbox{\boldmath $k$}_{\perp}'    -
                                 \mbox{\boldmath $k$}_{\perp}     +
                                 \mbox{\boldmath $\ell$}_{\perp} )
              \biggr ] \as (\lts) f(x, \lts)\, , \nonumber
\end{eqnarray}
where,
\begin{equation}\label{ddxg}
f(x,\ell^2)=\frac{\dd xG(x,\ell^2)}{\dd\ln\ell^2}
\end{equation}
(see \cite{LMRT} for details).
Substituting (\ref{tlamlam}) in (\ref{mlamlam1}) we have,
\begin{equation}\label{mlamlam2}
{\cal M}_\lamlam = i\frac{2\pi^2 s}{\sqrt{N_c}}\int\frac{d\lts}{\lt^4}
                  \as(\lts) f(x,\lts)\Delta\psi_\lamlam \, ,
\end{equation}where the definition of $\Delta\psi$ is to be understood from~
\eqs{mlamlam1}{mlamlam2}:
\begin{equation}\label{Deltapsi}
\Delta\psi = 2\psi(\mbox{\boldmath $k$}_{\perp},z) -
              \psi(\mbox{\boldmath $k$}_{\perp} - \mbox{\boldmath $\ell$}_{\perp},z) -
              \psi(\mbox{\boldmath $k$}_{\perp} + \mbox{\boldmath $\ell$}_{\perp},z)\, .
\end{equation}
$\psi_\lamlam $ has been calculated in
\cite{LMRT} both for a transverse and for a longitudinal  polarised incoming
photon. The wave function for a transverse polarized
photon and the amplitude for producing $\bar{q}q$ with helicities $\lambda$ and
$\lambda'$ read:
\newcommand{\lint}[2]{\int\frac{d #1^2}{#1^4}
                               \as(#2 ^2) f(x,#2 ^2)}
\newcommand{\lintarg}[3]{\left\{ \frac{1}{#3^2+#2^2} -
                                \frac{1}{\sqrt{\left( #3^2 +
                                                       #2^2 +
                                                      #1^2\right)^2-
                                                4 #2^2 #1^2}} \right\}}
\newcommand{\lintargb}[3]{\left\{ \frac{1}{#3^2+#2^2} -
                                  \frac{1}{2 #2^2} +
                            \frac{#3^2-#2^2+#1^2}{2 #2^2\sqrt{\left( #3^2 +
                                                       #2^2 +
                                                      #1^2\right)^2-
                                                4 #2^2 #1^2}} \right\}}
\newcommand{\lintargc}[3]{\left\{ \frac{#2^2 - #3^2}{#2^2 + #3^2 } +
                            \frac{#3^2-#2^2+#1^2}{\sqrt{\left( #3^2 +
                                                       #2^2 +
                                                      #1^2\right)^2-
                                                4 #2^2 #1^2}} \right\}}
\begin{eqnarray}
\psi^{\pm}_\lamlam\left({\mathbf \kt} ,z\right) 
   &=& -\delta_{\lambda , -\lambda'} Z_f e
                       \left[(1-2z)\lambda\mp 1\right] 
                 \frac{{2\mathbf\epsilon_{\pm}\cdot \kt}}{\bar{Q}^2+\kts} ,
                 \label{psilamlamt} \\
{\cal M}^{\pm}_\lamlam &=& 
 -2 Z_f e \left( i\frac{2\pi^2 s}{\sqrt{N_c}}\right)
 \delta_{\lambda , -\lambda'} \left[(1-2z)\lambda\mp 1\right] 
 {\mathbf\epsilon_{\pm}\cdot \kt}
                               \lint{\lt}{\lt}\,\times \nonumber \\
                              & & \hspace{1cm}\lintargb{\lt}{\kt}{\bar{Q}}
                               . \label{mlamlamt}
\end{eqnarray}
In (\ref{psilamlamt}) and (\ref{mlamlamt}), $Z_f$ is the charge of the quark
with flavour $f$ in units of the electron charge $-e$, $\bar{Q}^2\equiv
z(1-z)Q^2$ and $\epsilon_{\pm}$ denotes  the photon polarization 
vector 
presented in a circular basis,
{
\begin{equation}
\mathbf{\epsilon_{\pm}}=\frac{1}{\sqrt{2}}\left(0,0,1,\pm i\right).
\end{equation}

To evaluate the cross section one should, first, sum over the quark
helicities
$\lambda$ and $\lambda'$, and   average over the two transverse
polarisation
states $(\pm)$ of the photon
\begin{eqnarray}
|{\cal M}^T|^2 &=& \frac{32\pi^4 s^2}{N_c} Z^2_f e^2
                   \left[z^2+(1-z)^2\right]
                               \lint{\lt}{\lt}\,\times \nonumber \\
                            & & \hspace{1cm}\lintargc{\lt}{\kt}{\bar{Q}} .
\end{eqnarray}
The cross section is obtained by integration over $z$ and over $\kt$,
\begin{eqnarray}
\sigt &=&
\frac{\aem}{4\pi^2 N_c}\int^1_0 dz\left[z^2+(1-z)^2\right]
                       \int\frac{d \kts}{\bar{Q}^2+\kts}
                       \lint{\lt}{\lt}\,\times \nonumber \\
                         & & \hspace{3cm}\lintargc{\lt}{\kt}{\bar{Q}} .
\end{eqnarray}
Following \cite{GLMepj98}, in order to perform  the integration over
$z$, 
we introduce the variables $M$ and $\mt$:
\begin{eqnarray}
M^2 &=& \frac{\kts}{z(1-z)} \nonumber \\
\mt^2 &=& \frac{\lts}{z(1-z)} \label{mmtdef}\, ,
\end{eqnarray}
and rewrite the integrals of $\sigt$ in terms the variables $M$,
$\mathbf{\mt}$ and
{  $\mathbf{\lt}$,}
\newcommand{\ztola}{\sqrt{1-4\frac{\lts}{\mt^2}}}
\newcommand{\ztolb}{1-2\frac{\lts}{\mt^2}}
\begin{eqnarray}
\sigt &=& 
\frac{\aem}{4\pi^2 N_c}\int\frac{d M^2}{Q^2+M^2}\int\frac{d \mt^2}{\mt^2} 
\int\frac{d \lts}{\lts}\frac{\ztolb}{\ztola}
\as(\lts) f(x,\lts)\,\times \nonumber \\
                         & & \hspace{3cm}\lintargc{\mt}{M}{Q} .
\label{EQsight2}
\end{eqnarray}
Using a generalization
of the Gribov approach, we take a cutoff on the integration over $M^2$ and
insert the ratio (\ref{Rdef}) inside the integration sign,
\begin{eqnarray}
\sight &=& 
\frac{\aem}{2\pi^2 N_c}\int^{\infty}_{M^2_0}
\frac{d M^2\,\,R(M^2)}{Q^2+M^2}\int\frac{d \mt^2}{\mt^2} 
\int\frac{d \lts}{\lts}\frac{\ztolb}{\ztola}
\as(\lts) f(x,\lts)\,\times \nonumber \\
                         & & \hspace{1cm}\lintargc{\mt}{M}{Q} .
\label{EQsight3}
\end{eqnarray}
The last integral in (\ref{EQsight3}) can be integrated by parts, using
(\ref{ddxg}). In the limit $4\lts\ll\mt^2$, the value of the integral is
dominated by the upper limit of the integration, therefore we replace $\lts$
in $\as(\lts)$ and $xG(x,\lts)$ with $\mt^2/4$, and obtain, for three
colors
\begin{eqnarray}
\label{hardT}
\sight &=& \frac{2\pi\, \alpha_{em}}{3}
                  \int^{\infty}_{M^2_0}
                  \frac{d M^2 \,\,R(M^2)}{Q^2\,+\,M^2}\,\,
                  \int^{\infty}_{0}\frac{d{\tilde M}^2}{{\tilde M}^4}
                  \as(\frac{{\tilde M}^2}{4})\,x\,G(x,\frac{{\tilde M}^2}{4})
                  \times
\nonumber \\
& & \hspace{2cm}
\left\{\,\,\frac{M^2\,-\,Q^2}{M^2\,+\,Q^2}\,\,+\,\,\frac{Q^2\,+\,{\tilde
M}^2\,-\,M^2}
{\sqrt{(\,Q^2\,+\,M^2\,+\,{\tilde
M}^2\,)^2\,-\,4\,M^2\,{\tilde M}^2}}\,\,\right\}\,\,.
\end{eqnarray}
In the case of heavy quarks, we assume that the quark is heavy enough so that
 any contribution to the soft cross-section can be neglected. The heavy
 quarks contribution is then written with the replacements: $ 4\,M^2\,{\tilde
 M}^2 \to 4\,(M^2-4m_Q^2)\,{\tilde M}^2$, and $R(Q^2) \to
 R^{QQ}(M^2)$. $R^{QQ}$ is the heavy quark contribution to the ratio
 (\ref{Rdef}) and $m_Q$ is the mass of the heavy quark,
\begin{eqnarray}
\label{hardTQQ}
\sightQQ &=& \frac{2\pi\, \alpha_{em}}{3}
                  \int^{\infty}_{M^2_0}
                  \frac{d M^2 \,\,R^{QQ}(M^2)}{Q^2\,+\,M^2}\,\,
                  \int^{\infty}_{0}\frac{d{\tilde M}^2}{{\tilde M}^4}
                  \as(\frac{{\tilde M}^2}{4})\,x\,G(x,\frac{{\tilde M}^2}{4})
\nonumber \\
& & \mbox{\hfill}
\left\{\,\,\frac{M^2\,-\,Q^2}{M^2\,+\,Q^2}\,\,+\,\,\frac{Q^2\,+\,{\tilde
M}^2\,-\,M^2}
{\sqrt{(\,Q^2\,+\,M^2\,+\,{\tilde
M}^2\,)^2\,-\,4\,(M^2-4m_Q^2)\,{\tilde M}^2}}\,\,\right\}\,\,.
\end{eqnarray}
In the above formulae $x=x(M^2)=(Q^2+M^2)/W^2$ with $W$ being the
 center of mass  energy of the photon-nucleon interaction. 

\subsection{The Contribution of a Longitudinal Polarized Photon \label{secSIGL}}

\subsubsection{``Soft'' Contribution to $\sigma_L$ \label{subsecSIGSL}}
A priori, it is straight forward to write the formula for the ``soft''
 component of $\sigl$ in AQM. The result is similar to our model for $\sigt$,
 except that $M^2$ in the numerator should be replaced with the photon
 virtuality. This factor causes the longitudinal cross section to vanish for
 $Q^2 \to 0$ (see (\ref{alphapom}) for the values of the parameters).
\begin{equation}\label{softl}
   \sigsl = \softformula{Q^2}\left\{\dlformula\right\}.
\end{equation}

However, it turns out that the AQM over estimates the experimental data, and
 it has to be reduced with some phenomenological procedure, such as a
 numerical factor \cite{SchSpi} or a phenomenological function which
 decreases with $Q^2$~\cite{MRS}.  In our formalism the ratio between the AQM
 and pQCD contributions depends on the separation parameter $M_0$. Thus, by
 lowering the value of $M_0$ the ``soft'' component is suppressed. We have
 found that for the longitudinal part of the cross section, taking any value
 of $M_0$ which is below the resonances mass fits the experimental data, as
 opposed to the transverse cross section where we used $0.7 < M_0^2 < 0.9
 \gev2$ .

\subsubsection{``Hard'' Contribution to $\sigma_L$}
We calculate the pQCD contribution for $\sigl$, from the diagrams  of
Fig.~\ref{fig3}, using \eqs{mlamlam1}{Deltapsi}. For the case of a
longitudinal polarized incoming photon, we use the expression derived in 
\cite{LMRT}  for the wave function
\begin{equation}
\psi^{\scrbox{L}}_\lamlam = -2\delta_{\lambda , -\lambda'} Z_f e Q z(1-z)
                 \frac{1}{\bar{Q}^2+\kts} , \label{psiL}
\end{equation}
and substitute it in (\ref{mlamlam2}) to obtain the amplitude for a
longitudinal photon to produce $\bar{q}q$ with helicities $\lambda$ and
$\lambda'$ 
\begin{eqnarray}
{\cal M}_\lamlam^{\scrbox{L}} &=& -4\delta_{\lambda , -\lambda'} Z_f e Q z(1-z)
                               i\frac{2\pi^2 s}{\sqrt{N_c}}
                               \lint{\lt}{\lt}\,\times \nonumber \\
                              & & \hspace{4cm}\lintarg{\lt}{\kt}{\bar{Q}}
                               .
\label{mlamlaml}
\end{eqnarray}
The cross section is then obtained by squaring, summing over the final
helicities and integrating over $z$ and $\kt$,
\begin{eqnarray}
\sigl &=& \frac{16 \pi\aem}{N_c}\sum Z_f^2
                           Q^2 \int_0^1 dz \left[ z(1-z) \right]^2
                           \int \frac{d\kts}{\bar{Q}^2+\kts}
                           \lint{\lt}{\lt} \times \nonumber \\
                           & & \hspace{4cm}\lintarg{\lt}{\kt}{\bar{Q}}.
\label{EQsighl1}
\end{eqnarray}
Changing the  integration variables $\kts$ and $\lts$ to $M$ and
$\mt$,
respectively, using (\ref{mmtdef}), we have
\begin{eqnarray}
\sigl &=& \frac{16 \pi\aem}{N_c}\sum Z_f^2
           Q^2 \int_0^1 dz
               \int \frac{dM^2}{Q^2+M^2}
               \lint{\mt}{z(1-z)\mt} \times \nonumber \\
      & &      \hspace{4.5cm} \lintarg{\mt}{M}{Q} \, .
\label{EQsighl2}
\end{eqnarray}
We now write the formula for $\sighl$, in the same way that we did
in section \ref{secT}, by taking a cutoff on the integration over $M^2$, and
insert the ratio (\ref{Rdef}) inside the integration sign,
\begin{eqnarray}
\sighl &=& \frac{8 \pi\aem}{N_c}Q^2\int_{M_0^2}^\infty
           \frac{R(M^2)dM^2}{Q^2+M^2}\int_0^\infty\frac{d\mt^2}{\mt^4}
           \int_0^1 dz f\left(x,z(1-z)\mt^2\right)
                       \as\left(z(1-z)\mt^2\right) \times
           \nonumber \\
       & & \hspace{4.5cm}\lintarg{\mt}{M}{Q}\, .
\label{EQsighl3}
\end{eqnarray}
Recalling (\ref{ddxg}), we can integrate \eq{EQsighl3} by parts and obtain
(for $N_c=3$),
\newcommand{\msqrt}{\sqrt{(Q^2+M^2+\mt^2)^2-4 M^2 \mt^2}}
\begin{eqnarray}
\sighl &=& \frac{8\pi\aem}{3}Q^2\int_{M_0^2}^\infty
           \frac{R(M^2)dM^2}{Q^2+M^2}\int_0^\infty\frac{d\mt^2}{\mt^4}
           \, \overline{xG}\left(x,\mt^2\right)\times
           \nonumber \\
       & & \hspace{2cm}
           \left\{\frac{1}{Q^2+M^2} - \frac{1}{\msqrt} \,- \nonumber \right.\\
       & & \hspace{5.5cm}\left.
           \frac{\mt^2\left(Q^2+\mt^2-M^2\right)}
                {\left(\msqrt\right)^3}\right\},
\end{eqnarray}
where,
\begin{equation}
\overline{xG}\left(x,\mt^2\right) \equiv \int_0^1
                                        xG\left(x,z(1-z)\mt^2\right)
                                        \as\left(z(1-z)\mt^2\right) dz\, .
\end{equation}
For $\sighlQQ$, the contribution of heavy quarks to the longitudinal
component of $\sgp$, we replace
$ 4M^2{\tilde M}^2 $ by $4(M^2-4m_Q^2){\tilde M}^2$ and $R(Q^2)$ with
$R^{QQ}(M^2)$, as we did for the transverse part:
\newcommand{\msqrtQQ}{\sqrt{(Q^2+M^2+\mt^2)^2-4(M^2-4m_Q^2) \mt^2}}
\begin{eqnarray}
\sighlQQ &=& \frac{8\pi\aem}{3}Q^2\int_{M_0^2}^\infty
           \frac{R^{QQ}(M^2)dM^2}{Q^2+M^2}\int_0^\infty\frac{d\mt^2}{\mt^4}
           \, \overline{xG}\left(x,\mt^2\right)\times
           \nonumber \\
       & & \hspace{2cm}
           \left\{\frac{1}{Q^2+M^2} - \frac{1}{\msqrtQQ} \,- \nonumber \right.\\
       & & \hspace{4cm}\left.
           \frac{\mt^2\left(Q^2+\mt^2-M^2\right)}
                {\left(\msqrtQQ\right)^3}\right\}\, .
\label{hardLQQ}
\end{eqnarray}

\section{Comparison With The Experimental Data}
We now present the results for our calculation of $\sgp$ from the
 master equation
\begin{equation}\label{sigmatotal}
\sgp = \sigst + \sight + \sightQQ + \sigsl + \sighl + \sighlQQ \, .
\end{equation}
As stated, our model has three parameters, namely, $M_{0,L}, M_{0,T}$ and
 $\mu$. In \cite{GLMepj98} the relatively high value of $M_0=\sqrt{5}$ GeV
 was taken, in order to give an energy dependence that was in a reasonable
 agreement with the published experimental results. We found that if the hard
 longitudinal contributions are not neglected, the transverse separation
 parameter can be reduced to the value of $0.7 < M_{0,T}^2 < 0.9 \,\gev2$,
 while the longitudinal one can be taken to be $M_{0,L}^2 \lsim 0.4
 \,\gev2$. The results of the calculation are stable if the separation
 parameters are chosen within these bounds.

\begin{figure}[tbp]
\begin{center}
  \epsfig{file=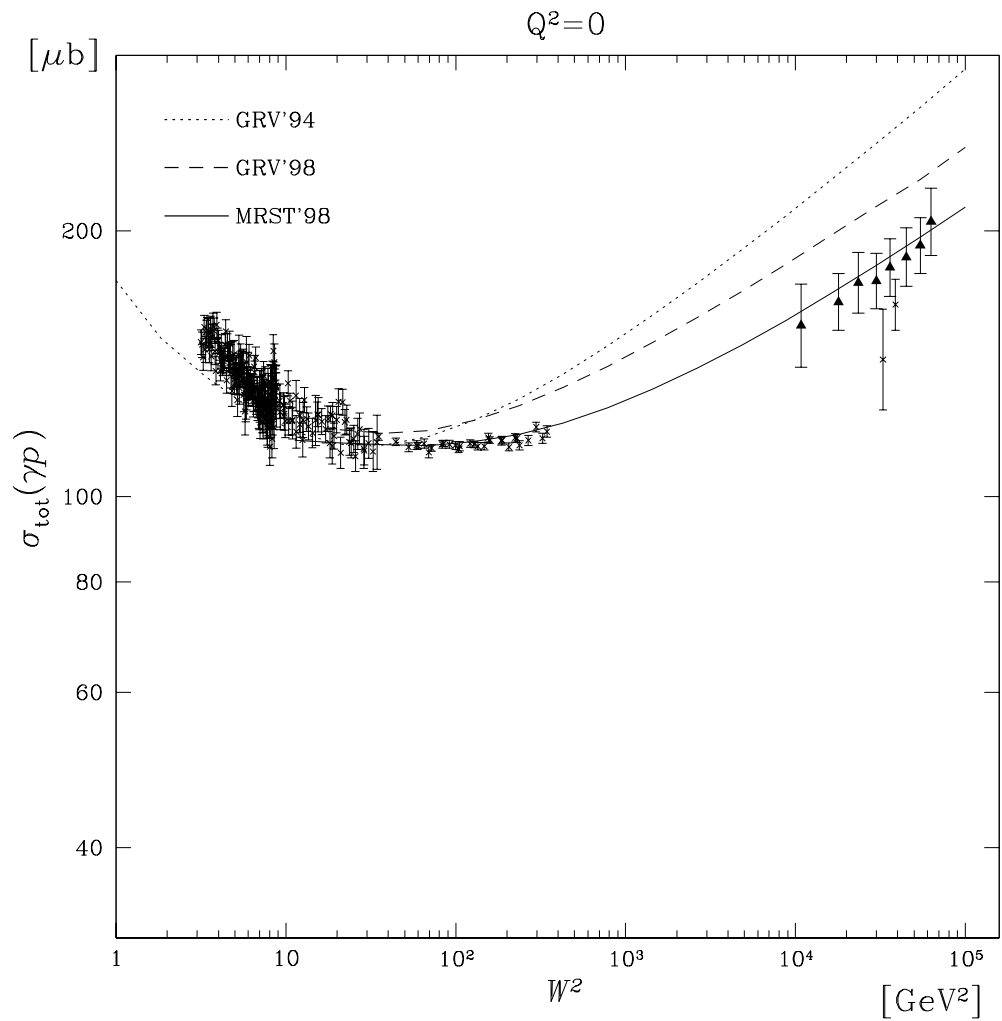,width=140mm}
  \caption[]{\parbox[t]{
             0.80\textwidth}{\small \it Comparison of  the gluon
             distribution
             parametrizaions which were used in Eq.\ \mbox{(\ref{sigmatotal})} 
             to calculate $\sgp$.  The full triangles correspond to data
             points
             which have been extracted in \cite{ZEUS98} from experimental 
              data.}
             \label{comparexg}}
\end{center}
\end{figure}

In the calculation of the hard components of $\sgp$, we need to specify the
 input gluon distribution $xG(x,Q^2)$ which appears in the formulae. We have
 tried several options: GRV'94 \cite{GRV94}, GRV'98 \cite{GRV98} and MRST'98
 \cite{MRST}. We compare the results of each input distribution for the
 calculation of $\sigma(\gamma p)$ in \fig{comparexg}, where it is obvious
 that the best parameterization for our purpose is MRST'98 which yields,
 together with our formalism, a good description of the data.

A difficulty in our calculation comes from the integration at very low
 $M^2$, where the published parameterizations for $xG$ are not valid.
 Naturally one has two possible options: the first is to impose a low cutoff
 on the integration below $M^2=\mu^2$. \,   The second option is to use
the general
 property of the gluon structure function which is linear in $Q^2$ at very
 small values of $Q^2$, and to rely on this property for the low mass region:
\begin{equation}\label{mu}
xG(x,l^2<\mu^2) = \frac{l^2}{\mu^2}xG(x,\mu^2)\, .
\end{equation}

In our calculation for the low mass integration we used~(\ref{mu}) for the
 gluon distribution. The coupling $\as(Q^2)$ was also kept fixed for
 $Q^2<\mu^2$. The value of $\mu^2$ was taken as the minimal value which is
 allowed in the input parameterization that had been used:
 $\mu^2=0.4,0.8,1.25\,\gev2$ for GRV'94, GRV'98 and MRST'98
 respectively. Notice that for the MRST parameterization $\mu^2=1.25\gev2 >
 M_0^2$. This means that the transition from soft to hard is not sharp, and
 we still have (\ref{mu}) as a soft signature inside our hard formalism at
 $M_0^2<M^2<\mu^2$.

\begin{figure}[tbp]
\begin{center}
  \epsfig{file=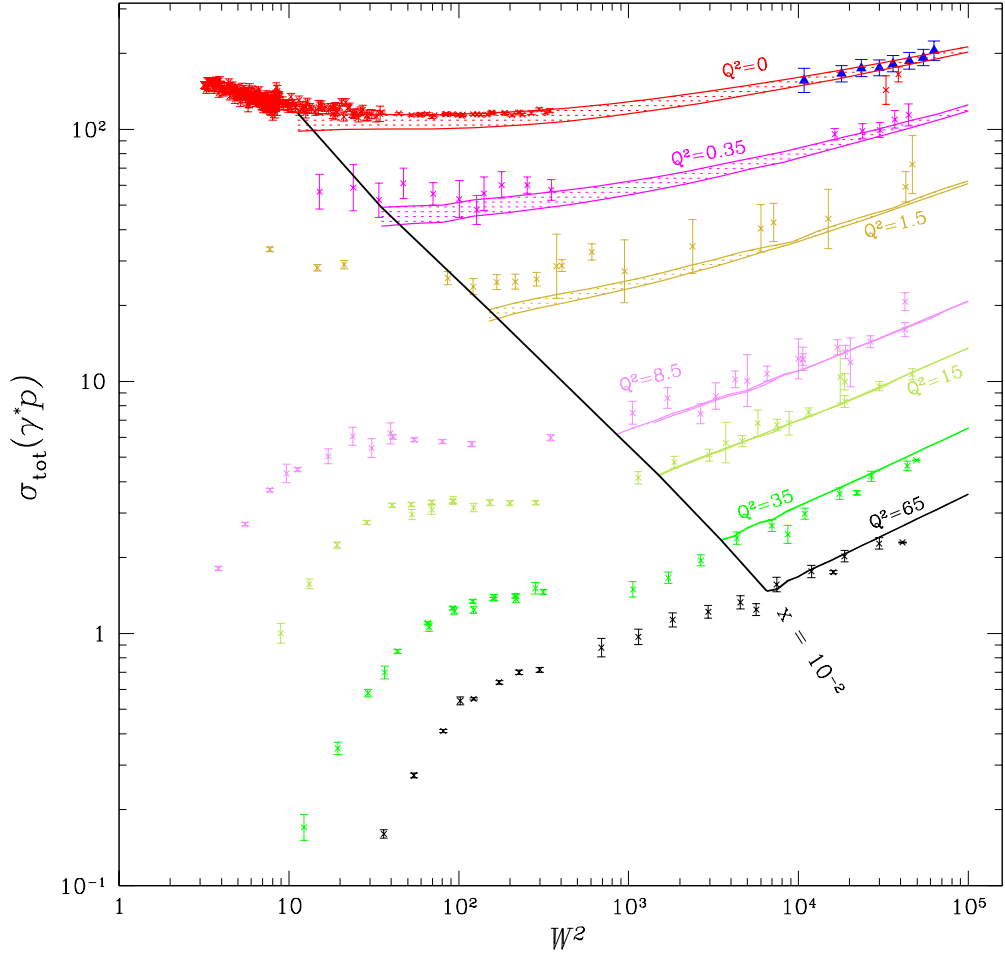,width=140mm}
  \caption[]{\parbox[t]{
             0.80\textwidth}{\small \it $\sgp$ as a function of $W^2$ together 
              with the 
             experimental data. The full triangles correspond to data points
             which have been extracted in \cite{ZEUS98} from experimental data.}
             \label{stotw2}}
\end{center}
\end{figure}
In \fig{stotw2} we show  the results of our calculation for $\sgp$ as
 a function of
 $W^2$ for fixed values of $Q^2$, together with the published experimental
 results. The  calculated results are shown as a band which
 corresponds to
 the two limits on $M_{0,T}$. It is clear from the figure that the width of
 the band decreases with increasing $Q^2$. As for the longitudinal separation
 parameter, the results are not sensitive to the choice of $M_{0,L}$, and in
 all our calculations we simply used $M_{0,L} = 2 m_\pi$.  The thick line
 marks the region of small $x$: to the left of it,  $x$ is not small
enough to
 justify using our model.  One can see that for $x < 10^{-2}$, our results
 reproduce the experimental results both in value and in energy dependence.

\begin{figure}[tbp]
\begin{center}
  \epsfig{file=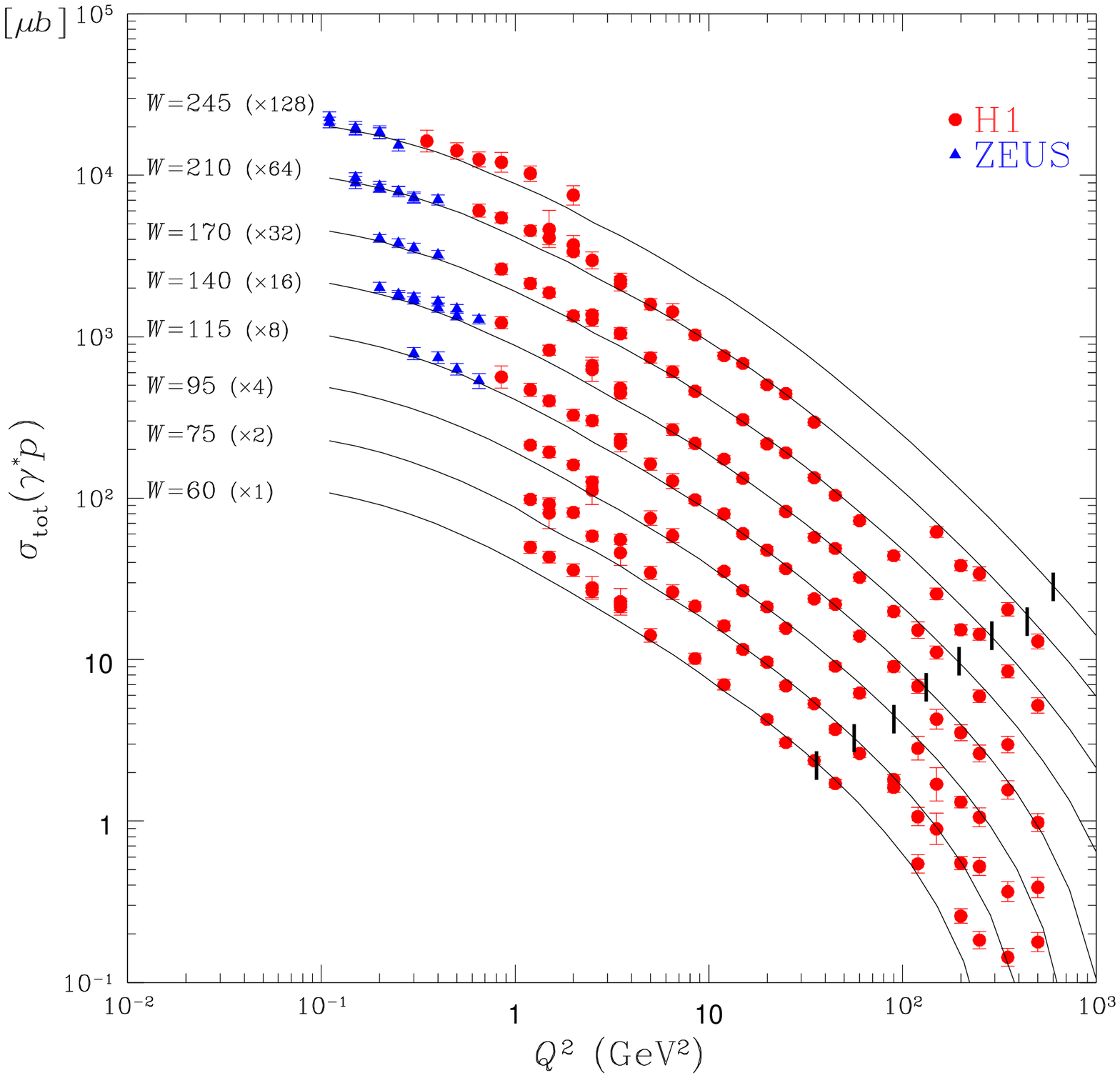,width=140mm}
  \caption[]{\parbox[t]{
             0.80\textwidth}{\small \it $\sgp$ as a function of $Q^2$ together 
              with the 
             experimental data.}
             \label{stotq2}}
\end{center}
\end{figure}

 In \fig{stotq2} we show our calculations for $\sgp$ for fixed values
of $W^2$,
as a
 function of $Q^2$, scaled by factors of $1-128$. The
small
 vertical thick lines delineate  the boundary $x = 10^{-2}$, to the
right of
 the marks our results are not reliable.

\begin{figure}[tbp]
\begin{center}
  \epsfig{file=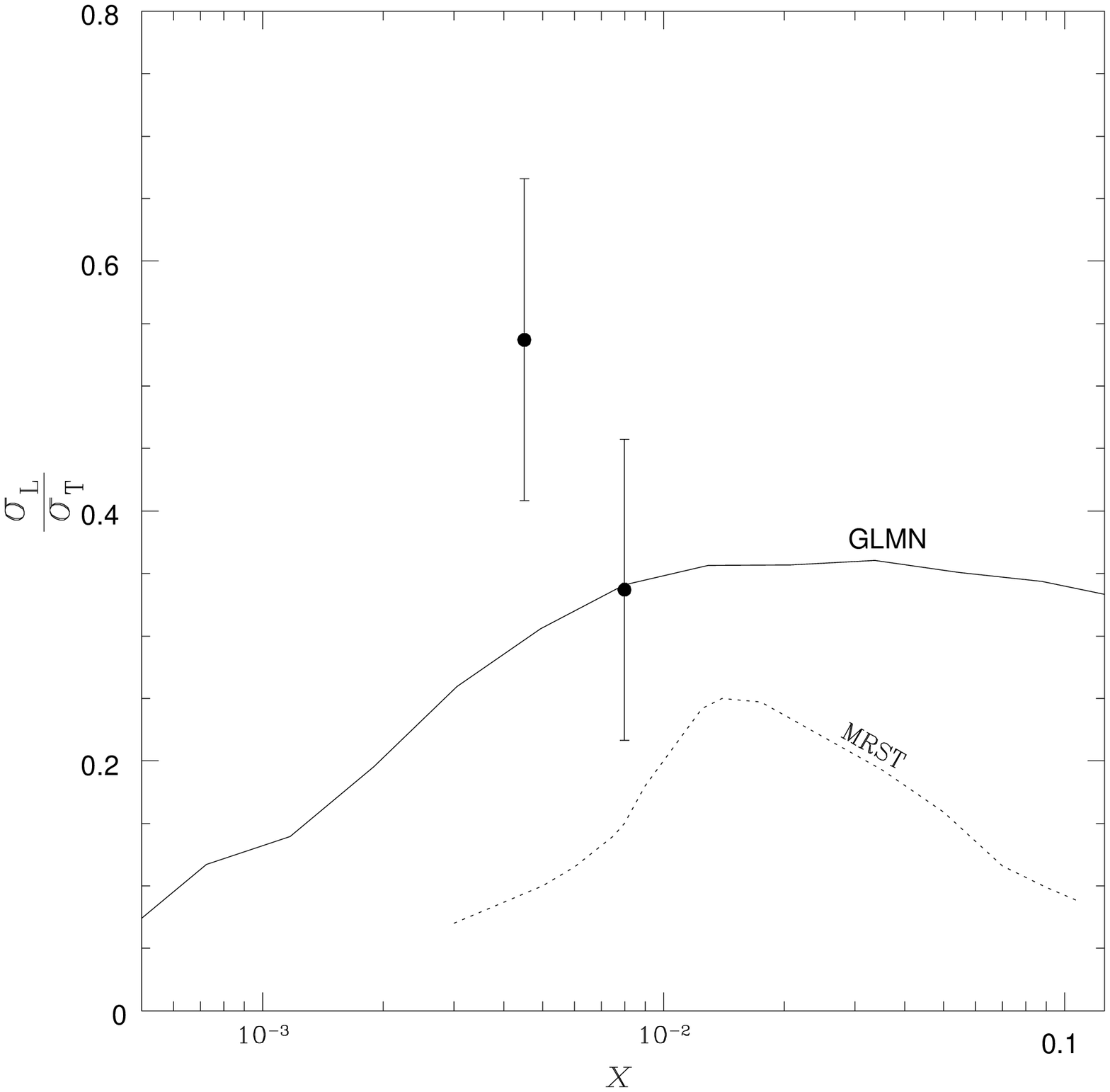,width=140mm}
  \caption[]{\parbox[t]{
             0.80\textwidth}{\small \it The ratio of $\ds{\frac{\sigl}{\sigt}}$ as
             a function of $x$.}
             \label{longtran}}
\end{center}
\end{figure}

The longitudinal component of $\sgp$, according to our calculation, reaches a
 maximum value of 25\precent of the total cross section at $x\approx
 10^{-2}$. For small $x$, this result is closer to the published data
 \cite{NMC} than the  results of Ref.\ \cite{MRS}, where they
 obtained $\sigl$ for
 $x\approx 10^{-2}$ to be 15\precent of the total cross section. The ratio
 of  the longitudinal cross section to the transverse cross section, as a
 function of $x$, is shown in \fig{longtran} together with the (small $x$)
 data of Ref.\ \cite{NMC}.

\begin{figure}[tbp]
\begin{center}
  \epsfig{file=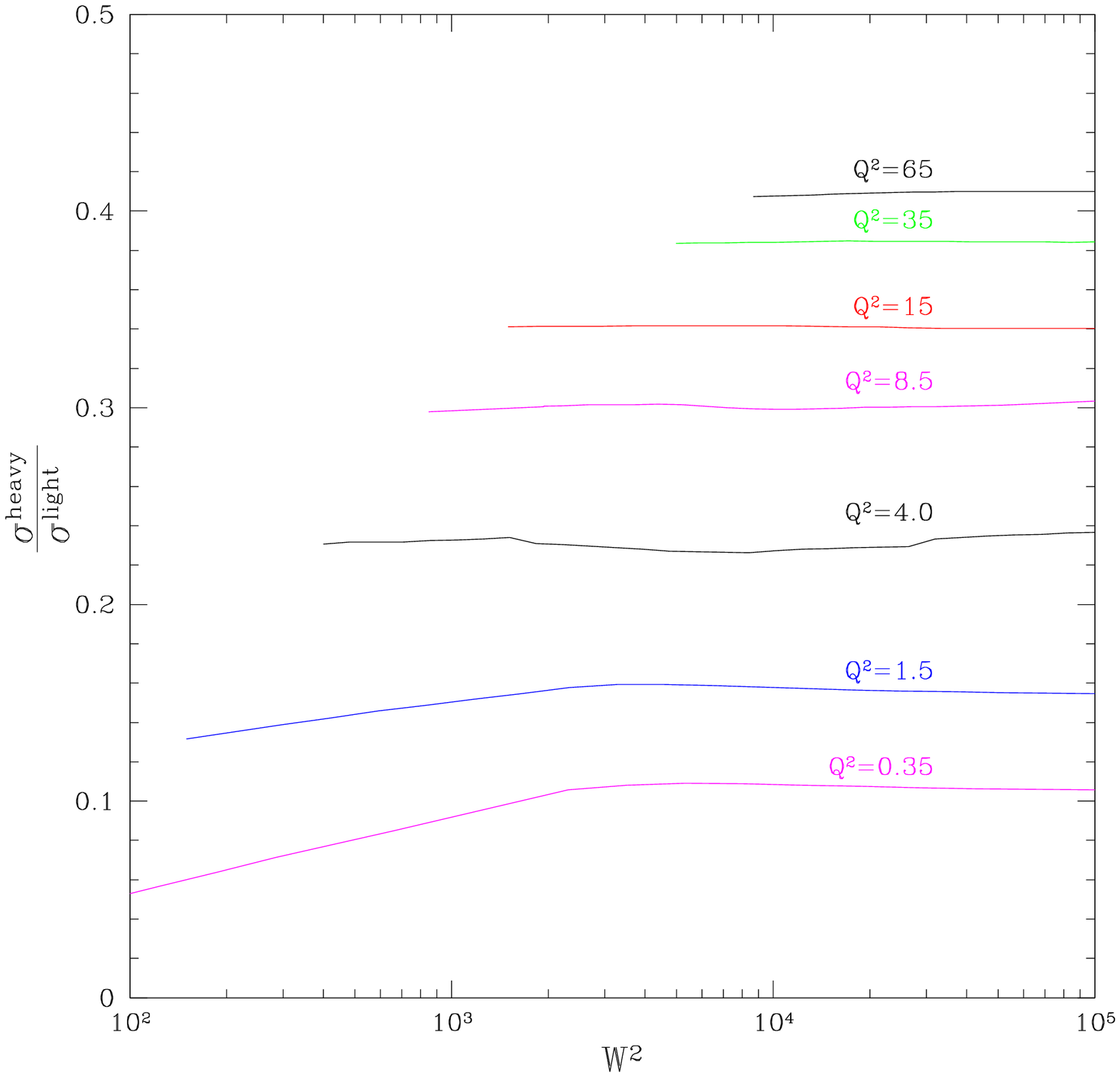,width=140mm}
  \caption[]{\parbox[t]{
             0.80\textwidth}{\small \it The ratio of
             $\frac{\sigma^{\scrbox{heavy}}}{\sigma^{\scrbox{light}}}$ at fixed
             $Q^2$.  }
             \label{heavylight}}
\end{center}
\end{figure}
\fig{heavylight} illustrates the importance of the heavy quarks contributions
 $\sightQQ$ and $\sighlQQ$ [see Eqs.\ (\ref{hardTQQ}) and
 (\ref{hardLQQ})]. The contribution has a mild $W^2$-dependence but it is
 $Q^2$-dependent, and for large $Q^2$, the ratio
 $\sigma^{\scrbox{heavy}}/\sigma^{\scrbox{light}}$ gets as high as $0.3 -
 0.4$ . Another ratio which we present is the ratio, of the hard to the
 soft contributions. Since $M_{0,L}$ is smaller than the lightest resonance
 mass, the AQM contribution to the longitudinal component is supressed, and
 therefore what we call ``soft'' contribution is actually the transverse AQM
 calculation.

\begin{figure}[tbp]
\begin{center}
  \epsfig{file=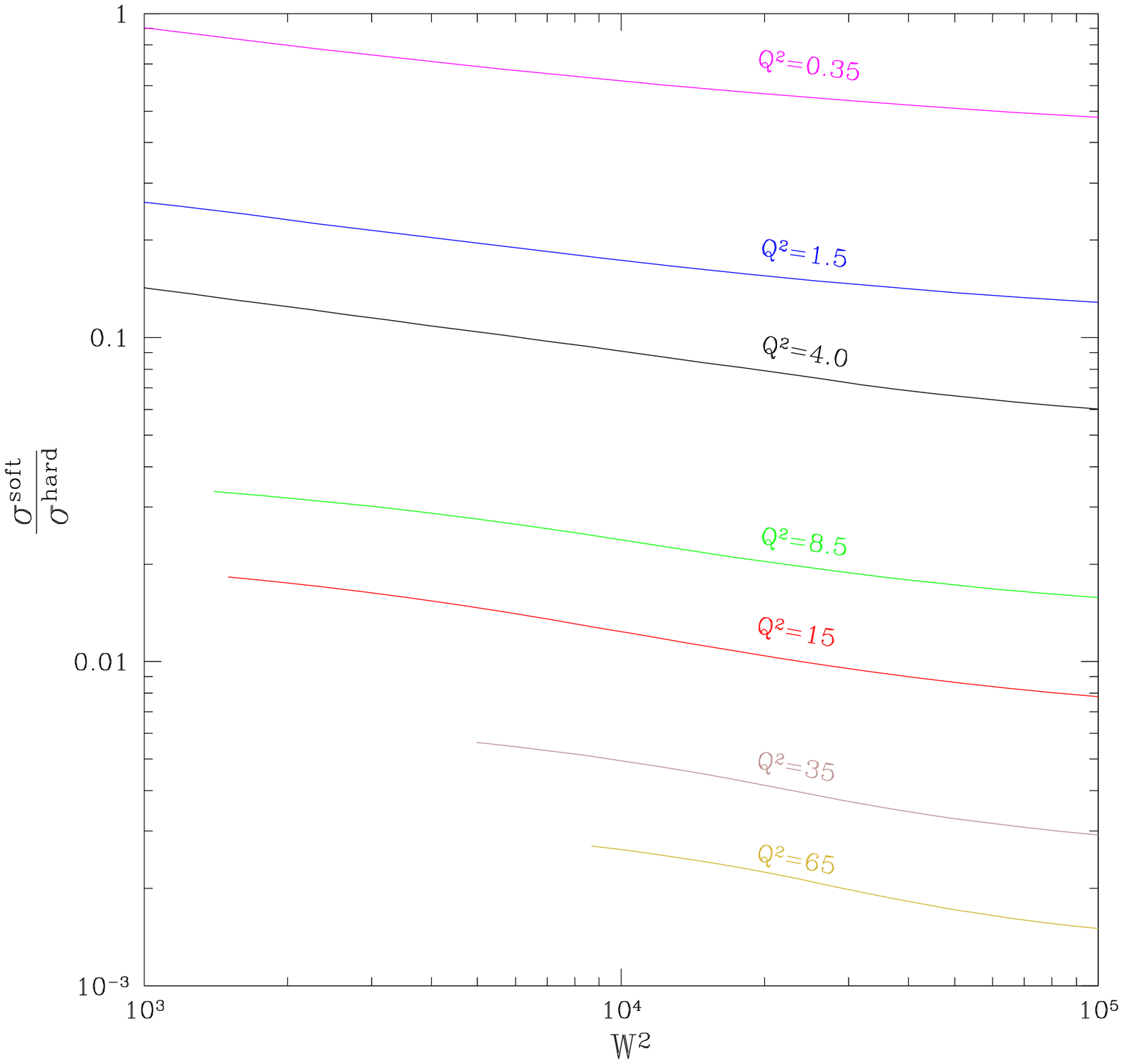,width=140mm}
  \caption[]{\parbox[t]{
             0.80\textwidth}{\small \it The ratio of
             $\frac{\sigma^{\scrbox{soft}}}{\sigma^{\scrbox{hard}}}$ at fixed
             $Q^2$.  }
             \label{softhard}}
\end{center}
\end{figure}

In \fig{softhard} we present the ratio
 $\sigma^{\scrbox{soft}}/\sigma^{\scrbox{hard}}$ at fixed values of $Q^2$, 
 where a clear power law behaviour as a function of $W^2$ can be seen. In
 soft processes, the hard contribution is minor, but is still present even for
 relatively small values of $Q^2$. The soft signature in hard processes is
 also present, but is smaller. For high virtualities it decreases from
 a few precent at low energy,  to less than 1\precent at high energies.

\section{Conclusions}

In this paper, we have shown that an approach based on Gribov's proposal, 
 provides a successful description of the experimental data on
photon-proton interaction, over a  wide range of photon virtualities $0\le
Q^2\le 100\,\gev2$, and energies $3\le\sqrt{s}\, (= W) \le 300$ GeV\@.  The
key assumption on which our aproach is based, is that the non-perturbative
and the perturbative QCD conributions in the Gribov formula can be separated
by the parameter $M_0$.  Oue successful reproduction of the experimental data
(see Figs.\ \ref{stotw2} and~\ref{stotq2}) shows that this assumption appears
to be valid. It further lends credence to using the additive quark model
(AQM) to describe the non-perturbative contribution.  The successful use of
the AQM leads to an improved result presented  in this paper, compared to
our previous
result\cite{GLMepj98}, where we found it necessary to introduce an damping
factor for  the AQM.

The second important by-product of our calculation is the simple method 
we have used to separate
between non-perturbative (``soft") and perturbative (``hard")
contributions. We have used three separation parameters:
$M_{0,T}$, $M_{0,L}$ and $\mu$. The values of these parameters, $M^2_{0,T} =
0.7 - 0.9 \,\gev2$, $M^2_{0,L}\leq 0.4\,\gev2 $ and $\mu^2\approx 1\,\gev2$, 
were determined by fitting to the experimental data.
We
believe that these values may be useful in the future,  for more
theoretical
description of the matching between ``soft" and ``hard" processes in QCD.

The third result which we find interesting, is that the GRV parameterizations
of the structure functions cannnot successfully describe the experimental
energy dependence of the total cross section at small values of $Q^2$ (see
\fig{comparexg}),  while the MRST parameterizations can.  This result
shows the
interdependence of deep inelastic scattering data, and the theoretical
description of the mathching of the ``soft" and ``hard" contributions.

In addition  we obtain:

\begin{enumerate}
\item 
 A description of the ratio $\sigl/\sigt$ (see \fig{longtran})
 which is in good agreement with the experimental data and other
 approaches. The fact that $\sigma_L$ appears to have only a ``hard"
 contribution should be stressed, as this could be a possible window for 
 particular features of non-pertutbative QCD;

\item 
 A considerable contribution of the ``soft" processes at rather large
 values of photon virtualities $Q^2$. For example at $Q^2 = 15\,\gev2$, $\sigs/\sigh
 \,\approx\,\,4\%$ at $W= 10 \,GeV$ (see \fig{softhard}). 
  We also find, at low $Q^2$, 
 a contamination of the ``soft" processes by the ``hard" ones. For high 
 energy real photoproduction this contamination amounts to 
 about 5\precent.
 We believe that this fact should be taken into account when
 interpreting experimental data, especially as far as their
 energy dependance is concerned;

\item 
 The prediction of the cross section for the heavy quark production (see
 \fig{heavylight}) which is almost $W$ independent, while showing a steep $Q^2$
 decrease.

\end{enumerate}

We propose a simple and successful 
phenomenological  model for the photon-proton interaction, which provides a 
method for matching the ``soft" and ``hard" interactions at high 
energy, and which can be a guide for future theoretical approaches. 
It is worthwhile mentioning, that even in the present form 
our model can be useful for determining the initial parton distributions at 
leading twist, for the DGLAP evolution equations. 

{\bf Acknowledgments:}

We thank A. Martin and A.  Stasto for useful discussions of the results of
this paper, and for providing us the kumac file for Fig.6. This research
was supporteed in part by the Israel Science Foundation, founded by the
Israel Academy of Science and Humanities.

\end{document}